\documentclass[11pt,a4paper]{article}
\usepackage[T1]{fontenc}

\usepackage[a4paper,top=2cm,bottom=2cm,left=3cm,right=3cm,marginparwidth=1.75cm]{geometry}
 
\usepackage{amsmath}
\usepackage{graphicx}
\usepackage{hyperref}
\usepackage{multirow}
\usepackage{lipsum}
\usepackage{amsmath}
\usepackage{amssymb}
\usepackage{graphicx}
\usepackage{float}
\usepackage{siunitx}
\usepackage{tikz}
\usetikzlibrary{arrows.meta, shapes.geometric}
\usetikzlibrary{calc}
\usepackage{caption}
\usepackage{subcaption}
\usepackage{graphicx} 
\usepackage{geometry}

\usepackage{algorithm}
\usepackage{algpseudocode}
\algdef{SE}[PARFOR]{ParFor}{EndParFor}
  {\textbf{parfor}}{\textbf{end parfor}}

\usepackage{xcolor}

\graphicspath{{./Figures/}}

\usepackage[affil-it]{authblk}

\makeatletter
\renewcommand{\section}{\@startsection {section}{1}{\z@}%
              {24pt}{12pt} {\large\scshape\bfseries}}

\renewcommand{\subsection}{\@startsection {subsection}{2}{\z@}%
             {12pt}{12pt}  {\itshape\bfseries}}

\usepackage{apacite}
\usepackage{natbib}

\title{\bfseries \normalsize Decentralized Model Predictive Control of Connected and Automated Vehicles with Coupled Safety Constraints}

\author{Philip Schultheis}
\author{Kimia Chavoshi*}
\author{John Lygeros}

\affil{ Automatic Control Laboratory, Department of Electrical Engineering and Information Technology,
ETH Zürich, Switzerland}

\date{\vspace{-5ex}}

\begin{document}

\maketitle

\section*{Abstract}\small
Connected and Automated Vehicles (CAVs) operating on lane-free highways offer substantial gains in traffic efficiency. However, their inherent nonlinear dynamics and the presence of coupled, nonconvex safety constraints present critical challenges to control design. Centralized Model Predictive Control (MPC) ensures safety, but suffers from scalability and communication limitations. To address these challenges, this paper investigates decentralized MPC (DMPC) for CAV coordination, focusing on iterative, non-cooperative algorithms, including Jacobi-type and Gauss-Seidel-type. A novel decoupling method is developed to transform nonconvex safety constraints into convex, locally enforceable constraints, inspired by buffered Voronoi cells. The simulation results show that the proposed DMPC algorithms achieve safe and efficient vehicle trajectories while substantially improving scalability, highlighting their potential for future lane-free CAV traffic systems. Ultimately, the results indicate that the most suitable decentralized control strategy depends on the desired trade-off between safety, performance, and computational efficiency.

\textbf{Keywords}: Connected and Automated Vehicles, Decentralized Control, Lane-Free Traffic.

\section{Introduction}
\label{sec:introduction}
    The rise of Connected and Automated Vehicles (CAVs) enables novel traffic paradigms, such as lane-free highways, where vehicles exploit the full capacity of the road to increase throughput and efficiency \cite{sekeran_lane_free_2022, Papageorgiou2024, chavoshi2025}. In this context, safety is paramount; vehicles must avoid collisions during complex lateral maneuvers. Model Predictive Control (MPC) is widely used here due to its prediction capabilities and constraint handling \cite{rawlings_mpc_2020}. While centralized MPC ensures safety, it scales poorly and imposes high communication demands \cite{christofides_distributed_2013, chavoshi_fairness_2024}.\\
    Decentralized MPC (DMPC) offers a scalable alternative by solving local subproblems in parallel or sequentially \cite{maestre2014dmpc}. However, coupled nonconvex constraints in DMPC can lead to compromised safety, infeasibility, and sub-optimality. To address these, we developed a novel method to decouple safety constraints.
    The main contributions of this work are:
    \begin{itemize}
        \item Developing an algorithm for decoupling nonconvex safety constraints.
        \item Developing a Jacobi-type algorithm to solve the DMPC problem with coupled and decoupled safety constraints.
        \item Developing a Gauss-Seidel-type algorithm to solve the DMPC problem with coupled and decoupled safety constraints.
        \item Analyzing safety, scalability, real-time feasibility, and performance of the proposed methods.
    \end{itemize}
    Section \ref{sec:related_work} reviews related work on lane-free CAV control and DMPC. Section \ref{sec:proposed_method} presents the problem formulation and the proposed decentralized algorithms, including the constraint decoupling method. Section \ref{sec:results} reports the simulation results. Finally, Section \ref{sec:discussion} discusses the main findings, and Section \ref{sec:summary} concludes the research and outlines future work.

    \section{Related Work}
	\label{sec:related_work}
    Lane-free traffic is an emerging paradigm in which CAVs are not restricted to discrete lanes, using methods such as nudging to optimize road capacity and improve traffic efficiency \cite{sekeran_lane_free_2022, Papageorgiou2024}. On lane-free highways, a kinematic bicycle model is often used to capture nonlinear lateral--longitudinal movement while remaining computationally tractable for receding-horizon control \cite{chavoshi_fairness_2024}, \cite{levy_path_2021}. Centralized nonlinear MPC formulations can incorporate such models and enforce collision avoidance constraints directly, but generally suffer from limited scalability and high communication requirements in dense traffic \cite{chavoshi_fairness_2024}.\\
    A common strategy to improve scalability is to formulate DMPC problems with linear (or linearized) kinematic bicycle dynamics and convex constraints, enabling efficient local solvers and clear convergence properties.
    Often, ADMM-based decomposition is used to solve linear DMPC with coupled constraints \cite{bai2023robust}, and consensus variants have been proposed to handle limited communication in large-scale motion planning \cite{lygerosADMM}, \cite{liu2024admm}. Additionally, \cite{dabestani2025distributed} addresses lane-free control using an event-triggered DMPC to reduce communication overhead, considering a linear double-integrator model.\\
    To retain maneuver realism (e.g., tight lateral maneuvers at highway speeds), several works consider nonlinear MPC formulations. A real-time NMPC framework based on convex--concave decomposition is presented in \cite{dong2024eco}, where nonconvex constraints are handled via iterative convex approximations. Related convexification ideas have been used in distributed CAV control implementations, e.g., via penalty-based convex--concave procedures (CCP) in experimental settings \cite{katriniok2022fully}. Furthermore, Jacobi-type iterative methods are a classical building block for DMPC of dynamically coupled systems and have been analyzed for stability and feasibility under convexity assumptions \cite{doan_jacobi_2008}.\\
    Maintaining nonlinear dynamics while avoiding overly restrictive convexifications is particularly important for dense lane-free scenarios where overtaking and strong lateral interactions occur. Recent nonlinear DMPC schemes include approaches based on optimality-condition decomposition (OCD) to coordinate vehicles without full centralization \cite{facerias_nonlinear_2024}, and sensitivity-based DMPC methods that reduce the computation per iteration by exploiting local sensitivity information under inexact solutions \cite{pierer2024sensitivity}. Although sensitivity-based and OCD methods address nonlinearity, they often struggle with coupled, nonconvex safety constraints caused by collision avoidance in lane-free traffic. These coupled, nonconvex safety constraints remain challenging to enforce in a fully decentralized architecture without sacrificing real-time feasibility. 
 
Unlike existing approaches that rely on centralized coordination or simplified linear models, our approach considers decentralized controllers and nonlinear dynamics and addresses the challenges posed by decoupled convex and coupled nonconvex safety constraints arising from collision avoidance.
    
    \section{Methodology}
	\label{sec:proposed_method}

    \subsection{Decentralized Model Predictive Control with Nonconvex Coupled Safety Constraints}\label{subsec:dmpc_coupled}
    Our starting point is the methodology proposed in \cite{chavoshi_fairness_2024}, in which a threat detection algorithm identifies CAV pairs that may possibly collide (threat relation) if not controlled appropriately. Afterwards, a threat cluster is introduced as a group of CAVs connected by threat relations, and a centralized MPC is developed to control CAV movement within each cluster. Here, we propose a non-cooperative iterative DMPC scheme to achieve scalability in dense lane-free traffic.
     Each CAV $i \in \{1, \dots, N\}$ solves a local Optimal Control Problem (OCP) based on its own state $\boldsymbol{s}_i(k)$ and the trajectories $(\boldsymbol{x}_{j}^{\text{a}}, \boldsymbol{y}_{j}^{\text{a}})$ announced by its threats $j \in \mathcal{T}_i$.\\
    The coupling between CAVs occurs primarily through collision-avoidance constraints. In a decentralized setting, a safety gap arises because the announced trajectory $(\boldsymbol{x}_{j}^{\text{a}}, \boldsymbol{y}_{j}^{\text{a}})$ used by CAV $i$ may differ from the actual trajectory $(\boldsymbol{x}_j^*, \boldsymbol{y}_j^*)$ simultaneously optimized by CAV $j$. This discrepancy can lead to constraints violations and infeasibility. To address this while maintaining the realism of lane-free maneuvers, we define the local OCP for CAV $i$ at time $t$ over a horizon $p$ as:
    \begin{subequations}\label{eq:DMPC_OCP}
    \begin{align}
        \min_{\boldsymbol{u_i}, \boldsymbol{\epsilon}} & \sum_{k=t+1}^{t+p} \Big( \|\boldsymbol{s}_i(k) - \boldsymbol{s}_i^{\text{d}}\|_Q^2 + \|\boldsymbol{u}_i(k)\|_R^2 + \sum_{j \in \mathcal{T}_i} \zeta \epsilon_j(k) \Big) \label{eq:cost_func} \\
        \mathrm{s.t.} \quad & \boldsymbol{s}_i(k+1) = f(\boldsymbol{s}_i(k), \boldsymbol{u}_i(k)), \label{eq:dynamics} \\
        & \boldsymbol{s}_i(k) \in \mathcal{S}_i, \quad \boldsymbol{u}_i(k) \in \mathcal{U}_i, \label{eq:state_input_bounds} \\
        & \mathcal{C}_{i,j}(\boldsymbol{s}_i(k), \boldsymbol{x}_{j}^{\text{a}}(k), \boldsymbol{y}_{j}^{\text{a}}(k),\epsilon_j(k)) \geq 0, \quad \forall j \in \mathcal{T}_i, \label{eq:coupled_cons}
    \end{align}
    \end{subequations}
    where $\boldsymbol{s}_i = [x_i, y_i, v_i, \theta_i, \delta_i]^\top$ and $\boldsymbol{u}_i = [a_i, \omega_i]^\top$ follow the nonlinear kinematic bicycle dynamics $f(\cdot)$
    
    \begin{equation}
    \label{eq: dynamics}
        \left\{
        \begin{aligned}
        x_i(k+1) &= x_i(k) + T v_i(k) \cos(\theta_i(k)) \\
        y_i(k+1) &= y_i(k) + T v_i(k) \sin(\theta_i(k)) \\
        v_i(k+1) &= v_i(k) + T a_i(k) \\
        \theta_i(k+1) &= \theta_i(k) + \tfrac{T}{l_i} v_i(k) \tan(\delta_i(k)) \\
        \delta_i(k+1) &= \delta_i(k) + T \omega_i(k).
        \end{aligned}
        \right.
    \end{equation}
    
    The cost function \eqref{eq:cost_func} penalizes deviations from the desired state $\boldsymbol{s}_i^{\text{d}}$ (including speed and heading angle) and control effort, while the final term introduces a slack variable $\epsilon_j$ to handle soft safety constraints. The parameters $Q$, $R$, and $\zeta$ are weights for different terms in the cost function, where $Q$ is positive semi-definite, $R$ is positive definite, and $\zeta$ is always positive. We assume that these parameters are identical for all CAVs. The physical limits on the values of the state variables and control signals are imposed by $\mathcal{S}_i$ and $\mathcal{U}_i$, respectively.
    
    The set of threat vehicles for CAV $i$ is denoted by $\mathcal{T}_i$. Inter-vehicle safety for threat pairs is enforced by nonconvex coupled constraints $\mathcal{C}_{i,j}$ in \eqref{eq:coupled_cons}, defined by safety and threat margins represented by ellipses centered at the centroid of vehicle $i$, with half-axis $r_{\text{ls}}$ and $r_{\text{ws}}$, for safety margin, and $r_{\text{lt}}$ and $r_{\text{wt}}$, for threat margin \cite{chavoshi_fairness_2024}. The threat margin is defined as the larger ellipse that encompasses the safety margin. Note that the major axis of the ellipses is aligned with the x-axis, and variations in the heading angle $\theta$ are not taken into account when defining their shape. We drop the prediction time step $k$ in the following to improve readability
    \begin{align}
        &\frac{(x_i - x_{j}^{\text{ann}})^2}{r_{\text{ls}}^2} + \frac{(y_i - y_{j}^{\text{ann}})^2}{r_{\text{ws}}^2} \geq 1 \label{eq:hard_safety} \\
        &\frac{(x_i - x_{j}^{\text{ann}})^2}{r_{\text{lt}}^2} + \frac{(y_i - y_{j}^{\text{ann}})^2}{r_{\text{ws}}^2} \geq 1 - \epsilon_j \label{eq:soft_safety}\\
        &\epsilon_j\geq 0,
    \end{align}
    where \eqref{eq:hard_safety} represents the hard collision avoidance constraint, which prohibits threat vehicles from entering the safety margin, and \eqref{eq:soft_safety} is a soft threat-detection constraint that penalizes proximity to threat margin \cite{chavoshi_fairness_2024} for every prediction time step $k$. 
    
    Solving \eqref{eq:DMPC_OCP} directly is challenging due to the combination of non-convex equality and inequality constraints. In the following sections, we describe how we regain feasibility and safety through geometric constraint decoupling (Section \ref{subsec:decoupling}) and iterative solving (Section \ref{subsec:solvers}).

    \begin{figure*}[t]
        \centering
        \tikzset{
            egoStyle/.style={ultra thick, solid, black},
            threatStyle/.style={thick, dashed, black},
            tangentStyle/.style={thick, densely dotted, gray}
        }
    
        \begin{subfigure}{0.32\textwidth}
            \centering
            \resizebox{\textwidth}{!}{%
            \begin{tikzpicture}[scale=0.3]
                \def\carL{5} \def\carW{2} \def\ella{11} \def\ellb{3}
                \coordinate (C1) at (0,0); \coordinate (C2) at (-15,10);
                
                \begin{scope}[shift={(C1)}]
                    \draw[egoStyle, fill=gray!20] (-\carL/2, -\carW/2) rectangle (\carL/2, \carW/2);
                    \draw[egoStyle] (0,0) ellipse ({\ella} and {\ellb});
                    \coordinate (Pint) at (-4.2, 2.8); 
                    \draw[ultra thick, black] ($(Pint)+(-0.8,-0.8)$) -- ($(Pint)+(0.8,0.8)$);
                    \draw[ultra thick, black] ($(Pint)+(-0.8,0.8)$) -- ($(Pint)+(0.8,-0.8)$);
                    \node[above right, inner sep=4pt] at (Pint) {\Huge $\mathbf{P_{i,int}}$};
                \end{scope}
                
                \begin{scope}[shift={(C2)}]
                    \draw[threatStyle, fill=gray!5] (-\carL/2, -\carW/2) rectangle (\carL/2, \carW/2);
                    \draw[threatStyle] (0,0) ellipse ({\ella} and {\ellb});
                \end{scope}
                
                \fill[black] (C1) circle (0.2); \fill[black] (C2) circle (0.1);
                \draw[black, dash dot, thin] (C1) -- (C2);
            \end{tikzpicture}
            }
            \caption{Two vehicles with safety ellipses and centroid connection line.}
            \label{fig:vehicles_ell}
        \end{subfigure}
        \hfill
        \begin{subfigure}{0.32\textwidth}
            \centering
            \resizebox{\textwidth}{!}{%
            \begin{tikzpicture}[scale=0.3]
                \def\carL{5} \def\carW{2} \def\ella{11} \def\ellb{3}
                \coordinate (C1) at (0,0); \coordinate (C2) at (-15,10);
                
                \begin{scope}[shift={(C1)}]
                    \draw[egoStyle, fill=gray!20] (-\carL/2, -\carW/2) rectangle (\carL/2, \carW/2);
                    \draw[egoStyle] (0,0) ellipse ({\ella} and {\ellb});
                \end{scope}
                \begin{scope}[shift={(C2)}]
                    \draw[threatStyle, fill=gray!5] (-\carL/2, -\carW/2) rectangle (\carL/2, \carW/2);
                    \draw[threatStyle] (0,0) ellipse ({\ella} and {\ellb});
                \end{scope}
                
                
                \draw[egoStyle] (-20.3, 1.94) -- (9.48, 5.27); 
                \draw[threatStyle] (-24.4, 4.72) -- (5.32, 8.05); 
                
                \fill[black] (C1) circle (0.2); \fill[black] (C2) circle (0.1);
                
                \coordinate (Pides) at (-5.4, 3.6); 
                \draw[ultra thick, black] ($(Pides)+(-0.8,-0.8)$) -- ($(Pides)+(0.8,0.8)$);
                \draw[ultra thick, black] ($(Pides)+(-0.8,0.8)$) -- ($(Pides)+(0.8,-0.8)$);
                \node[above right, inner sep=4pt] at (Pides) {\Huge $\mathbf{p_{i,\text{des}}}$};
    
                \coordinate (Pjdes) at (-9.6, 6.4); 
                \draw[ultra thick, black] ($(Pjdes)+(-0.8,-0.8)$) -- ($(Pjdes)+(0.8,0.8)$);
                \draw[ultra thick, black] ($(Pjdes)+(-0.8,0.8)$) -- ($(Pjdes)+(0.8,-0.8)$);
                \node[below left, inner sep=4pt] at (Pjdes) {\Huge $\mathbf{p_{j,\text{des}}}$};

                \draw[black, dash dot, thin] (C1) -- (C2);
            \end{tikzpicture}
            }
            \caption{Construction of shifted half-space constraints from ellipse tangents.}
            \label{fig:vehicles_ell_tang}
        \end{subfigure}
        \hfill
        \begin{subfigure}{0.32\textwidth}
            \centering
            \resizebox{\textwidth}{!}{%
            \begin{tikzpicture}[scale=0.3]
                \def\carL{5} \def\carW{2} \def\ella{11} \def\ellb{3}
                \coordinate (C1) at (-5.4, 3.6); \coordinate (C2) at (-9.6, 6.4);
                \coordinate (Co1) at (0,0); \coordinate (Co2) at (-15,10);
                
                \begin{scope}[shift={(Co1)}]
                    \draw[egoStyle, opacity=0.6] (-\carL/2, -\carW/2) rectangle (\carL/2, \carW/2);
                    \draw[egoStyle, opacity=0.0] (0,0) ellipse ({\ella} and {\ellb});
                \end{scope}
                \begin{scope}[shift={(Co2)}]
                    \draw[threatStyle, opacity=0.6] (-\carL/2, -\carW/2) rectangle (\carL/2, \carW/2);
                    \draw[threatStyle, opacity=0.0] (0,0) ellipse ({\ella} and {\ellb});
                \end{scope}
                
                \begin{scope}[shift={(C1)}]
                    \draw[egoStyle, fill=gray!30] (-\carL/2, -\carW/2) rectangle (\carL/2, \carW/2);
                    \draw[egoStyle] (0,0) ellipse ({\ella} and {\ellb});
                \end{scope}
                \begin{scope}[shift={(C2)}]
                    \draw[threatStyle, fill=gray!5] (-\carL/2, -\carW/2) rectangle (\carL/2, \carW/2);
                    \draw[threatStyle] (0,0) ellipse ({\ella} and {\ellb});
                \end{scope}
                
                \draw[egoStyle] (-20.3, 1.94) -- (9.48, 5.27);
                \draw[threatStyle] (-24.4, 4.72) -- (5.32, 8.05);
            \end{tikzpicture}
            }
            \caption{Worst-case scenario: vehicles at their respective decoupled boundaries.}
            \label{fig:vehicles_ell_final}
        \end{subfigure}
        \caption{Geometric safety constraint decoupling process. Ellipses depict safety margins as in \eqref{eq:hard_safety}. Solid lines correspond to vehicle $i$ and dashed lines correspond to vehicle $j$.}
        \label{fig:geometric_decoupling_full}
    \end{figure*}
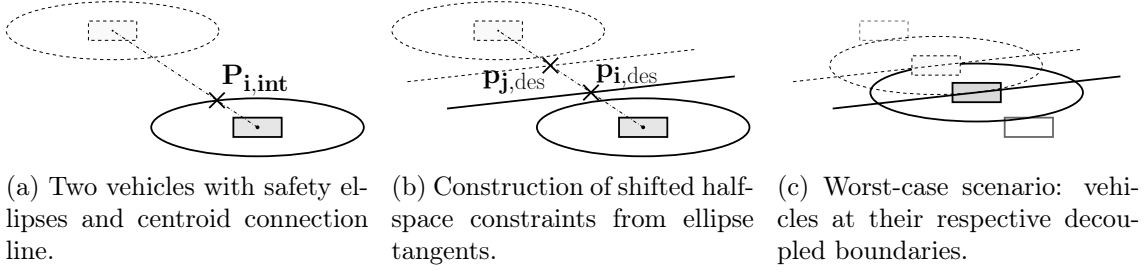


    
    \subsection{Geometric Safety Constraint Decoupling}\label{subsec:decoupling}
    To address the nonconvexity of the coupled safety constraints \eqref{eq:hard_safety}-\eqref{eq:soft_safety}, we propose a geometric decoupling algorithm inspired by Buffered Voronoi Cells (BVC) \cite{voronoi_graphic}. This method transforms the inter-vehicle safety requirements into local, convex half-space constraints.
    The algorithm builds safe regions around each vehicle that the centroid of the vehicle should not leave during a specific prediction step. In contrast to that, the initial safety ellipses (\ref{eq:hard_safety})-(\ref{eq:soft_safety}) determine areas where threats are not allowed to enter (coupled safety constraints). As illustrated in Fig. \ref{fig:vehicles_ell}, for any threat pair $(i, j)$, we consider the safety ellipses and the line connecting their centroids. 
    The decoupling is achieved by constructing a tangent line to the safety ellipse at the point where it intersects the centroid connection line as shown in Fig. \ref{fig:vehicles_ell_tang}. These tangents are shifted to their respective desired points $\mathbf{p_{i,\text{des}}}$ and $\mathbf{p_{j,\text{des}}}$ to obtain the lines in Fig. \ref{fig:vehicles_ell_tang}. The distance between $\mathbf{p_{i,\text{des}}}$ and $\mathbf{p_{j,\text{des}}}$ is equal to the distance between $\mathbf{P_{i,int}}$ and the centroid of the lower vehicle (Fig. \ref{fig:vehicles_ell}). This ensures that even if both vehicles move to the edge of their respective constraint (Fig. \ref{fig:vehicles_ell_final}), their centroids never enter the safety ellipse of the respective threatening vehicle.    
    The resulting convex polytope for each vehicle $i$ is defined by the intersection of half-spaces $\vec{n}_j^\top [x_i, y_i]^\top \leq \vec{n}_j \cdot \vec{p}_{i,\text{des}}$ for all threats $j$, where $\vec{n}_j$ is normal to the tangent and $\vec{p}_{i,\text{des}}$ is the shifted feasible boundary. Thus, we obtain individual convex polytopes for each prediction time step $k$. 
    
    \begin{algorithm}[h]
        \caption{Geometric Safety Constraint Decoupling for prediction time step $k$ from the perspective of vehicle $i$ to vehicle $j$.}
        \label{algo:decoupling}
        \begin{algorithmic}[1]
        \Require Predicted trajectories $\boldsymbol{s}_i, \boldsymbol{s}_j^{\text{ann}}$ and $r_{\text{ls}}, r_{\text{ws}}$
        \Ensure Half-space constraint $\mathcal{H}_{ij}$
        \State Identify $\mathbf{P_{i,int}}$.
        \State Compute the normal vector $\vec{n}_{ij}$ of the tangent line of the ellipse at $\mathbf{P_{i,int}}$.
        \State Compute $\mathbf{p_{i,\text{des}}}$ based on the distance between $\mathbf{P_{i,int}}$ and centroid $i$. 
        \State Shift tangent to $\mathbf{p_{i,\text{des}}}$ to obtain the solid black line in Fig. \ref{fig:vehicles_ell_tang} with normal vector $\vec{n}_{ij}$.
        \State Define half-space $\mathcal{H}_{ij}$: $\vec{n}_{ij}^{\top}[x_{i},y_{i}]^{\top} \le \vec{n}_{ij} \cdot \vec{p}_{i,\text{des}}$.
        \State \Return $\mathcal{H}_{ij}$
        \end{algorithmic}
    \end{algorithm}

    \subsection{Decentralized Iterative Solvers}\label{subsec:solvers}
    
    To solve the motion planning problem, we evaluate two iterative, non-cooperative DMPC algorithms that handle the nonlinear dynamics while incorporating the constraints described above. 

    \subsubsection{Jacobi Algorithm (Synchronous Updates)}
    The Jacobi scheme (Algorithm \ref{algo:jacobi}) leverages parallelization by allowing all CAVs to solve their local OCPs simultaneously. In each iteration $\text{I}$, every CAV $i$ optimizes its trajectory based on the trajectories announced by the other members of the threat cluster in the previous iteration $\text{I}-1$. This structure is ideal for high-speed parallel computation. 
    \begin{algorithm} 
    \caption{Jacobi DMPC (Synchronous)}
    \label{algo:jacobi}
    \begin{algorithmic}[1]
    \Require Warm-start trajectories $\boldsymbol{s}_{i, \text{warm}}$
    \Ensure Optimal control sequence $\boldsymbol{u}_i^*$
    \State Initialize $\boldsymbol{s}_{i}^{\text{ann}} \gets \boldsymbol{s}_{i, \text{warm}}$ 
    \For{$\text{I} = 1$ \textbf{to} $\text{I}_{\text{max}}$}
        \ParFor{ all CAVs $i \in \{1, \dots, N\}$}
            \State \textbf{4a. Coupled}: Receive $\boldsymbol{s}_{j}^{\text{ann}}$ from threats 
            \Statex \hspace{2.5em} $j \in \mathcal{T}_i$ and impose  \eqref{eq:hard_safety} and \eqref{eq:soft_safety}
            \Statex \hspace{2.5em} \textbf{4b. Decoupled}: Receive $\boldsymbol{s}_{j}^{\text{ann}}$ from $j \in$
            \Statex \hspace{2.5em} $\{1, \dots, N\}\setminus i$ and compute half-spaces 
            \Statex \hspace{2.5em} $\mathcal{H}_{ij}$ based on $\boldsymbol{s}_{j}^\text{ann}$ and $\boldsymbol{s}_{i}^\text{ann}$
            \State Solve local OCP \eqref{eq:DMPC_OCP} to obtain $(\boldsymbol{s}_{i}^{\text{I}}, \boldsymbol{u}_{i}^{\text{I}})$
        \EndParFor
        \State $\boldsymbol{s}_{i}^{\text{ann}} \gets \boldsymbol{s}_{i}^{\text{I}}$ for all $i$ \Comment{Synchronous update}
    \EndFor
    \State \Return $\boldsymbol{u}_i^*$
    \end{algorithmic}
    \end{algorithm}

    \subsubsection{Gauss-Seidel Algorithm (Asynchronous Updates)}
    The Gauss-Seidel (GS) approach (Algorithm \ref{algo:gauss_seidel}) introduces a sequential update order in which CAVs solve their local OCPs. This allows CAV $i$ to utilize the already optimized trajectories of CAVs $1, \dots, i-1$ within the current iteration $\text{I}$. Although the GS scheme limits parallelization, it typically achieves faster convergence, which has been also validated but is not the focus of this paper.
    \begin{algorithm}
    \caption{Gauss-Seidel DMPC (Asynchronous)}
    \label{algo:gauss_seidel}
    \begin{algorithmic}[1]
    \Require Warm-start trajectories $\boldsymbol{s}_{i, \text{warm}}$, priority ordering $\mathcal{P}$
    \Ensure Optimal control sequence $\boldsymbol{u}_i^*$
    \State Initialize $\boldsymbol{s}_{i}^{\text{ann}} \gets \boldsymbol{s}_{i, \text{warm}}$ 
    \For{$\text{I} = 1$ \textbf{to} $\text{I}_{\text{max}}$}
        \For{each CAV $i$ in order $\mathcal{P}$}
            \State \textbf{4a. Coupled}: Receive $\boldsymbol{s}_{j}^{\text{ann}}$ from threats 
            \Statex \hspace{2.5em} $j \in \mathcal{T}_i$ and impose \eqref{eq:hard_safety} and \eqref{eq:soft_safety}
            \Statex \hspace{2.5em} \textbf{4b. Decoupled}: Receive $\boldsymbol{s}_{j}^{\text{ann}}$ from $j \in$
            \Statex \hspace{2.5em} $\{1, \dots, N\}\setminus i$ and compute half-spaces 
            \Statex \hspace{2.5em} $\mathcal{H}_{ij}$ based on $\boldsymbol{s}_{j}^\text{ann}$ and $\boldsymbol{s}_{i}^\text{ann}$
            \State Solve local OCP \eqref{eq:DMPC_OCP} to obtain $(\boldsymbol{s}_{i}^{\text{I}}$, $\boldsymbol{u}_{i}^{\text{I}})$
            \State $\boldsymbol{s}_{i}^{\text{ann}} \gets \boldsymbol{s}_{i}^{\text{I}}$ \Comment{Asynchronous update}
        \EndFor
    \EndFor
    \State \Return $\boldsymbol{u}_i^*$
    \end{algorithmic}
    \end{algorithm}

    \section{Results}
	\label{sec:results}
	To evaluate the proposed DMPC schemes, we conduct extensive simulations in high-density, lane-free highway scenarios. We compare the Jacobi and GS solvers under two constraint regimes: the original nonconvex coupled constraints and the proposed convex decoupled half-spaces. The analysis focuses on safety guarantees, computational scalability, and tracking performance. For ease of notation, we introduce the abbreviations J\textsubscript{C}, J\textsubscript{D}, GS\textsubscript{C}, and GS\textsubscript{D} for the coupled and decoupled versions of Jacobi and Gauss-Seidel algorithms.

    \subsection{Simulation Setup}
    The simulation environment is based on the lane-free highway setup described in \cite{chavoshi_fairness_2024}, which involves 10 homogeneous CAVs on a 13.5\, m wide road. The fleet consists of five leading CAVs with a desired velocity of 80\,km/h and five following CAVs with a desired velocity of 100\,km/h. All the CAVs have the initial velocity of 80\,km/h.  We consider two traffic densities: an uncongested scenario (initial time gap 1\,s) and a congested scenario (0.75\,s). For each scenario, we conducted 10 case studies with different initial positions of CAVs. All simulations and control computations were executed in a MATLAB 2024b environment using the IPOPT solver on a workstation equipped with an Intel Core i7-13700K CPU (3.4 GHz) and 32 GB of RAM. We run Algorithms \ref{algo:jacobi} and \ref{algo:gauss_seidel} with a fixed maximum number of iterations $\text{I}_{\text{max}}$ of 5.
    

    \begin{table}[htbp]
        \centering
        \caption{Simulation parameter values \cite{chavoshi_fairness_2024}.}
        \label{tab:params}
        \small 
        \begin{tabular*}{\columnwidth}{@{\extracolsep{\fill}}lcc@{\extracolsep{\fill}}}
        \hline
        \textbf{Parameter} & \textbf{Symbol} & \textbf{Value} \\ 
        \hline
        Sim. / Control time step & $T, T_c$ & 0.05 s, 0.25 s \\
        Predict. / Time horizon & $p, T_A$ & 16, 4 s \\
        Road segment length & $L$ & 2500 m \\
        CAV size ($l \times w$) & $l_i, w_i$ & $5 \times 2$ m \\
        Safety ellipse axes & $r_{\text{ls}}, r_{\text{ws}}$ & 11 m, 3 m \\
        Threat ellipse axes & $r_{\text{lt}}, r_{\text{wt}}$ & 15 m, 3.2 m \\
        State (angle and speed) / Input weight & $Q(\alpha \hspace{1pt}\textrm{and}\hspace{1pt} \beta), R$ & (0.25 and 0.001), 0.0001 \\
        Soft constraint penalty weight& $\zeta$ & $10^3$ \\
        Minimum / Maximum velocity & $v_{\min}, v_{\max}$ & 0 km/h, 120 km/h \\
        
        Minimum / Maximum acceleration & $a_{\min}, a_{\max}$ & $-10.92$ m/s$^2$, $5.72$ m/s$^2$ \\
        Max. accel. change & $\Delta a$ & 0.7 m/s$^2$ \\
        Heading / Steering bound & $\Delta \theta$, $\Delta \delta$ & $\pi/3$ rad, $\pi/6$ rad\\
        Angular velocity bound & $\Delta \omega$ & $2\pi/3$ rad/s \\
        \hline
        \end{tabular*}
    \end{table}

	\subsection{Safety}\label{subsec:safety}
    \begin{figure}[htbp] 
        \centering
        
        \begin{subfigure}[b]{0.48
        \columnwidth}
            \centering
            \includegraphics[width=\linewidth]{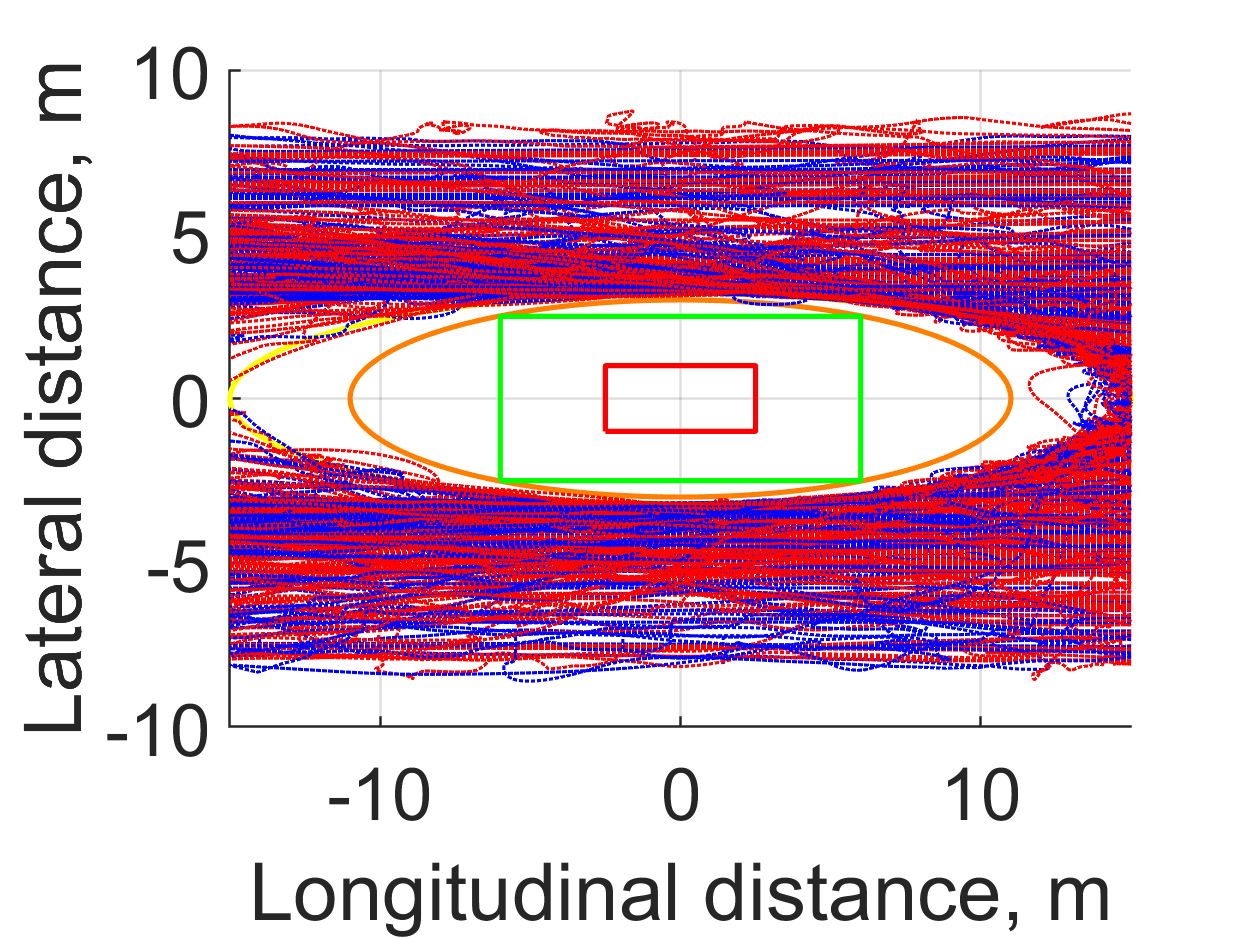}
            \caption{J\textsubscript{C}}
            \label{fig:safe_ell_rec_j_c}
        \end{subfigure}
        \hfill
        \begin{subfigure}[b]{0.48\columnwidth}
            \centering
            \includegraphics[width=\linewidth]{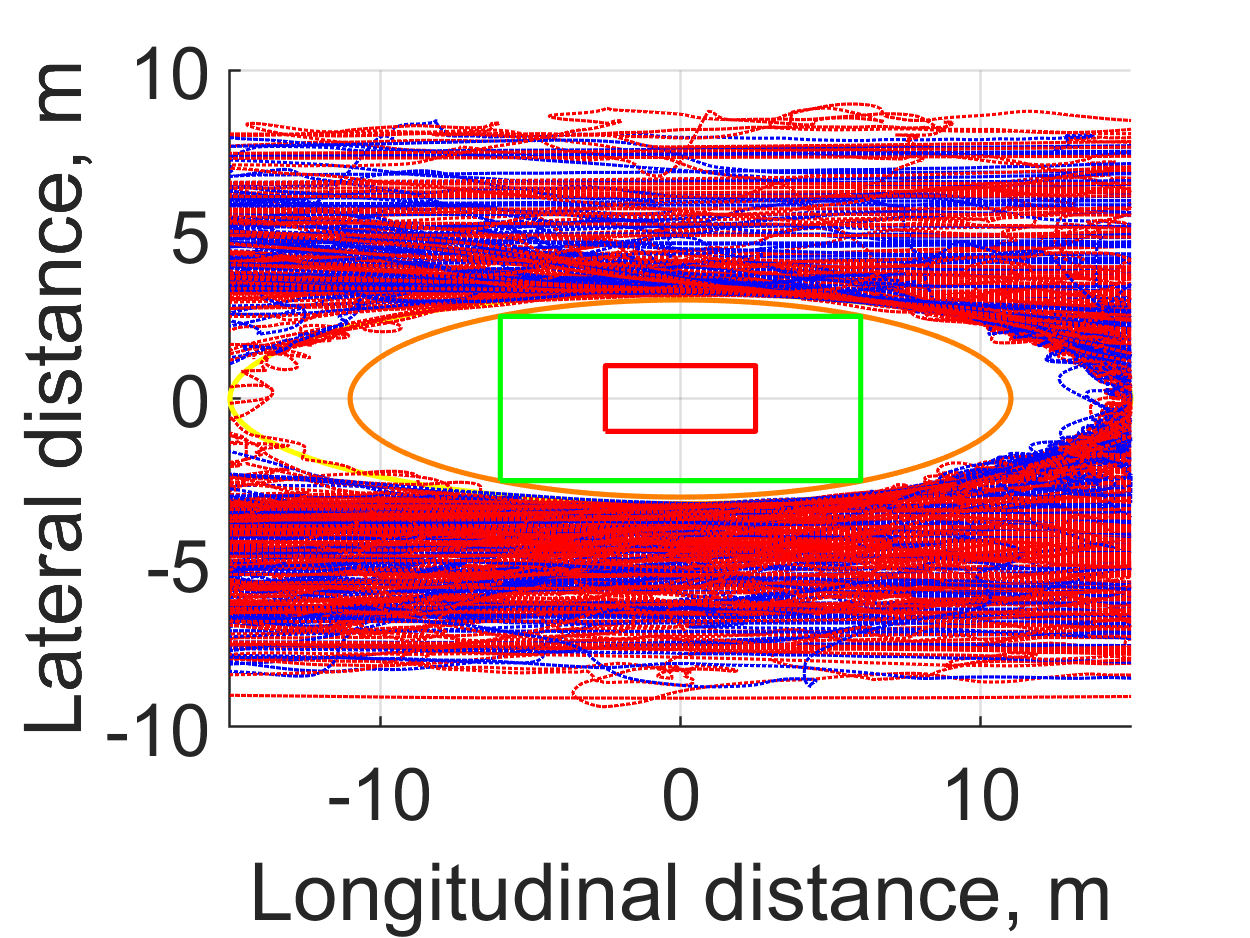}
            \caption{GS\textsubscript{C}}
            \label{fig:safe_ell_rec_gs_c}
        \end{subfigure}
    
        \vspace{4mm} 
    
        \begin{subfigure}[b]{0.48\columnwidth}
            \centering
            \includegraphics[width=\linewidth]{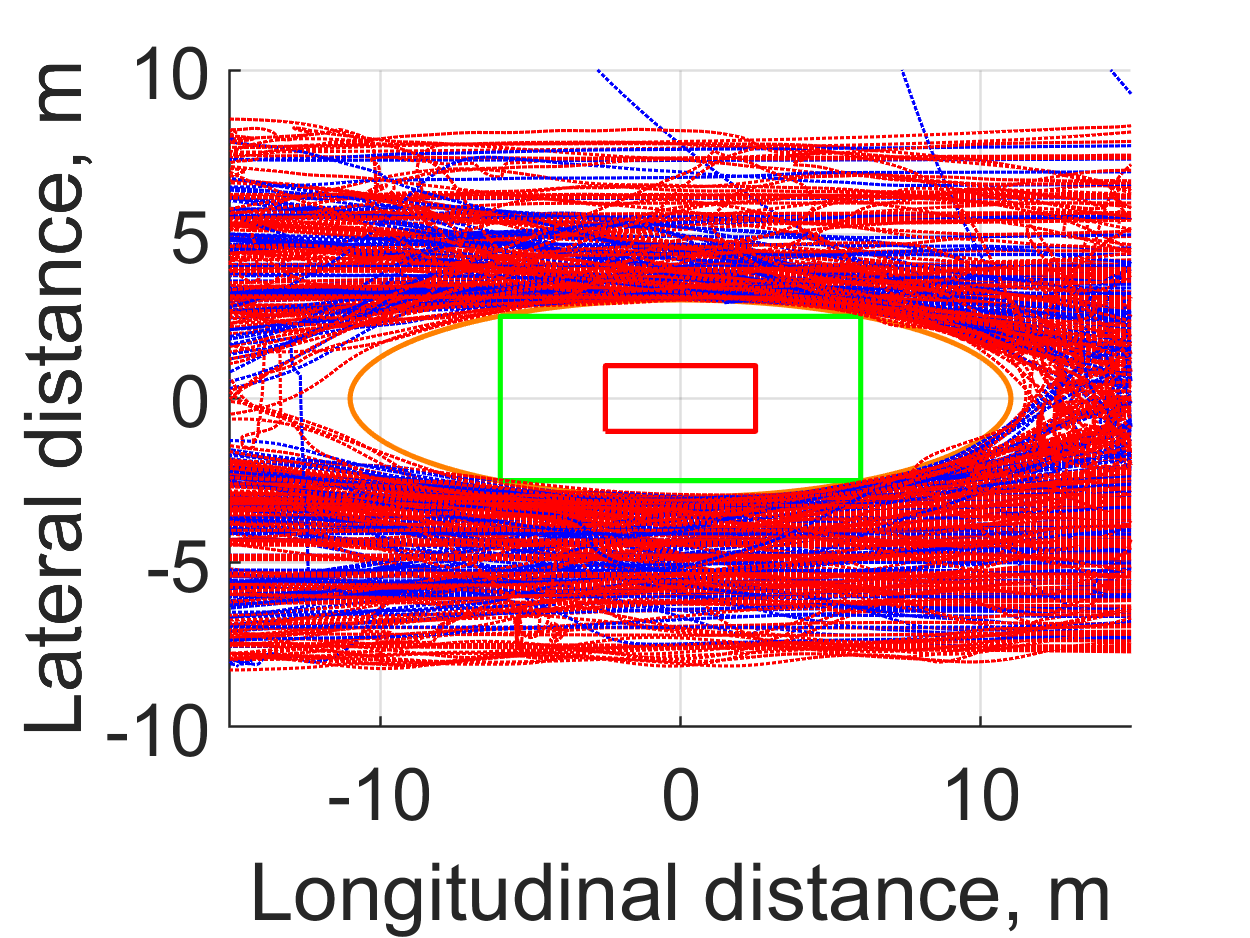}
            \caption{J\textsubscript{D}}
            \label{fig:safe_ell_rec_j_d}
        \end{subfigure}
        \hfill
        \begin{subfigure}[b]{0.48\columnwidth}
            \centering
            \includegraphics[width=\linewidth]{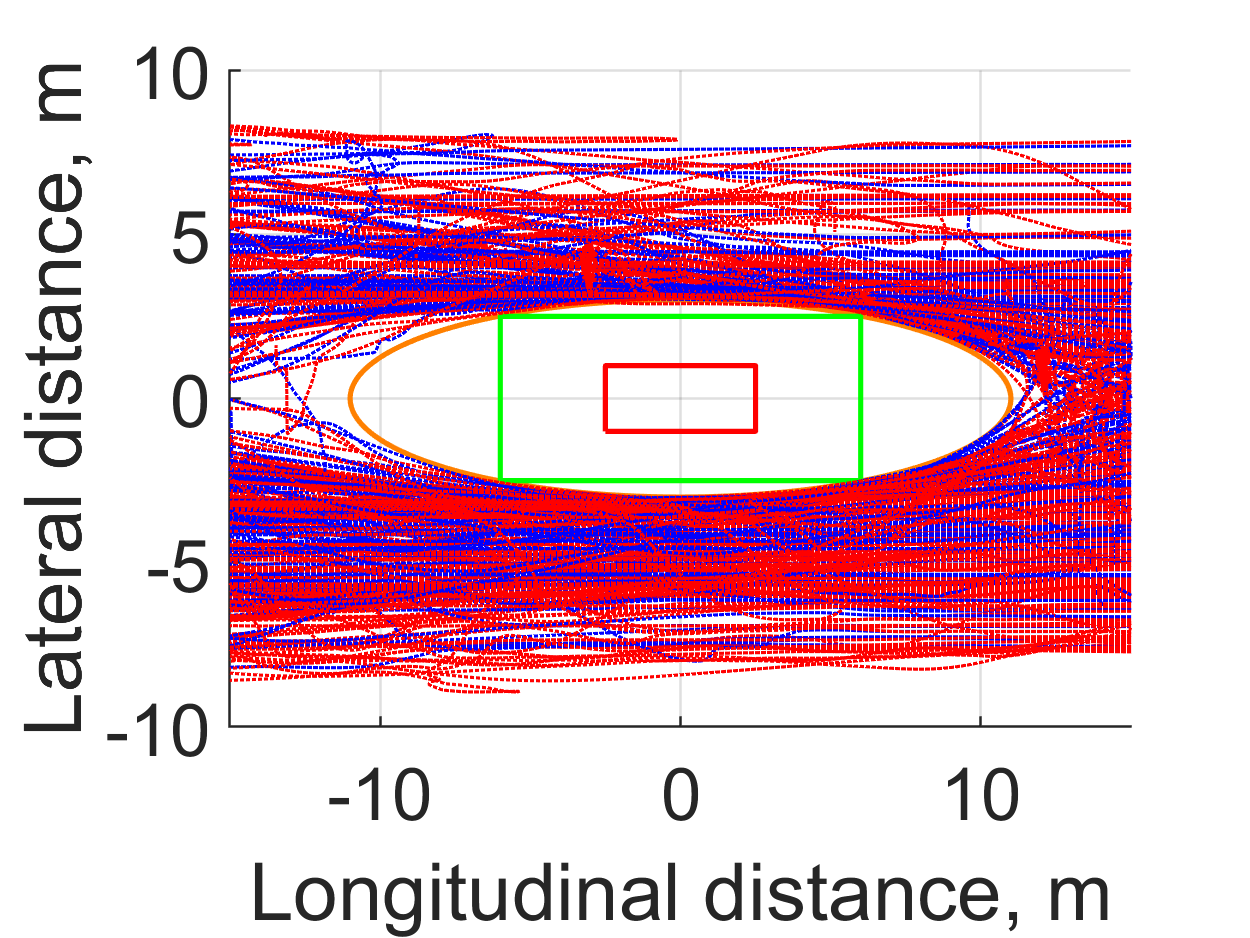}
            \caption{GS\textsubscript{D}}
            \label{fig:safe_ell_rec_gs_d}
        \end{subfigure}
    
        \caption{Safety illustration: CAV (red), collision zone (green), safety margin (orange), threat margin (yellow), trajectories of other CAVs (dotted lines).}
    \label{fig:safe_ell_rec}
    \end{figure}

    Safety is assessed by verifying that inter-vehicle distances remain outside the defined safety margins for all simulated instances.  Figure \ref{fig:safe_ell_rec} is a relative spatial illustration of (\ref{eq:hard_safety}) and (\ref{eq:soft_safety}), which is why both axes are in meters and indicate the lateral and longitudinal distances from the reference CAV’s centroid. If there was a trajectory (one of the dotted lines) entering the collision zone (green rectangle), it would indicate a collision. Both Jacobi and GS solvers successfully maintain safety (no collisions) after five iterations, demonstrating the robustness of the decentralized schemes. While the coupled solvers handle the original nonlinear ellipses, the decoupled variants utilize convex half-spaces that restrict the feasible region to a safe polytope. 
    
    \subsection{Scalability and Real-Time Feasibility}\label{subsec:scala}
    
    \begin{figure}[htbp] 
        \centering
        
        \begin{subfigure}[b]{0.48\columnwidth}
            \centering
            \includegraphics[width=\linewidth]{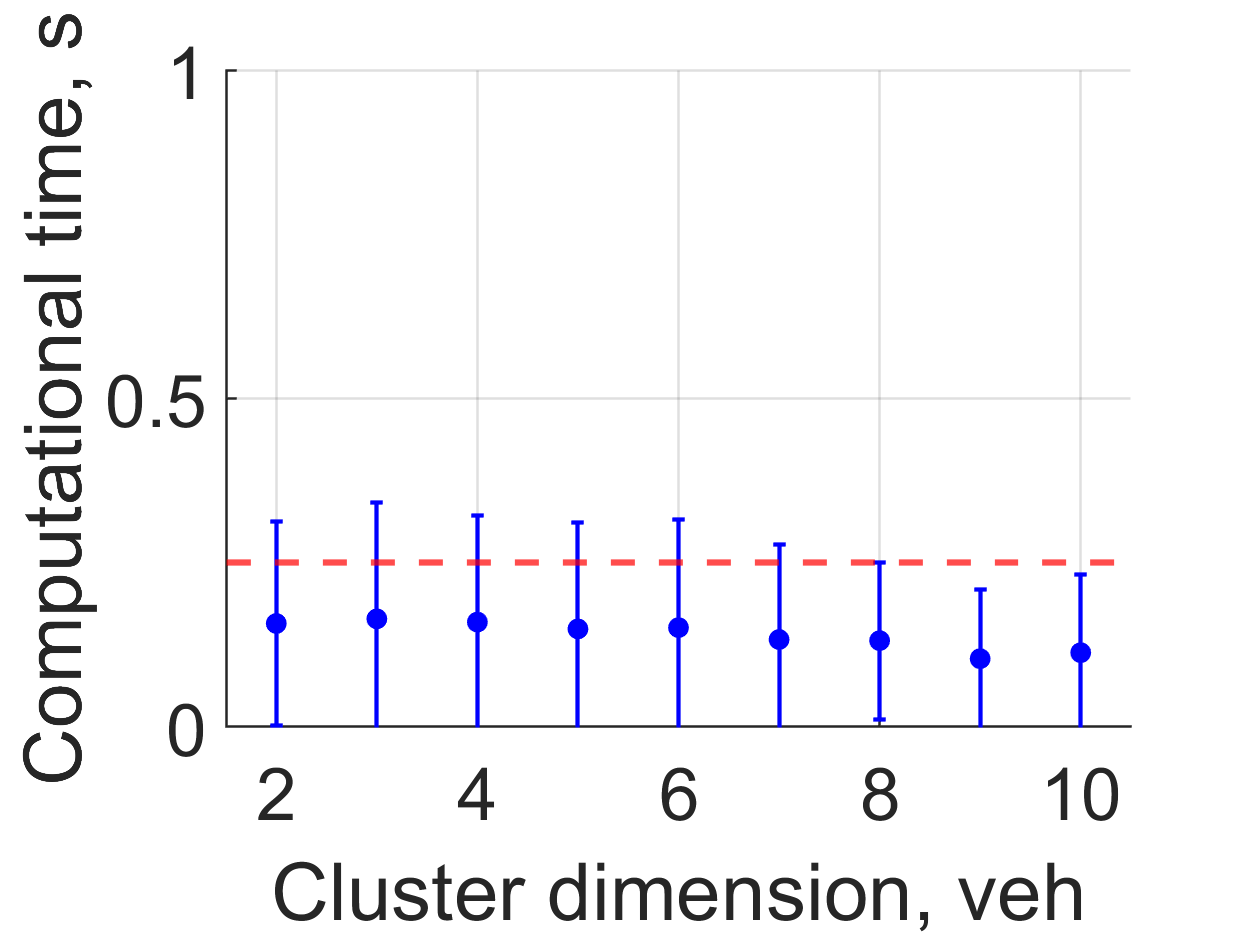}
            \caption{J\textsubscript{C}}
            \label{fig:ct_j_c}
        \end{subfigure}
        \hfill
        \begin{subfigure}[b]{0.48\columnwidth}
            \centering
            \includegraphics[width=\linewidth]{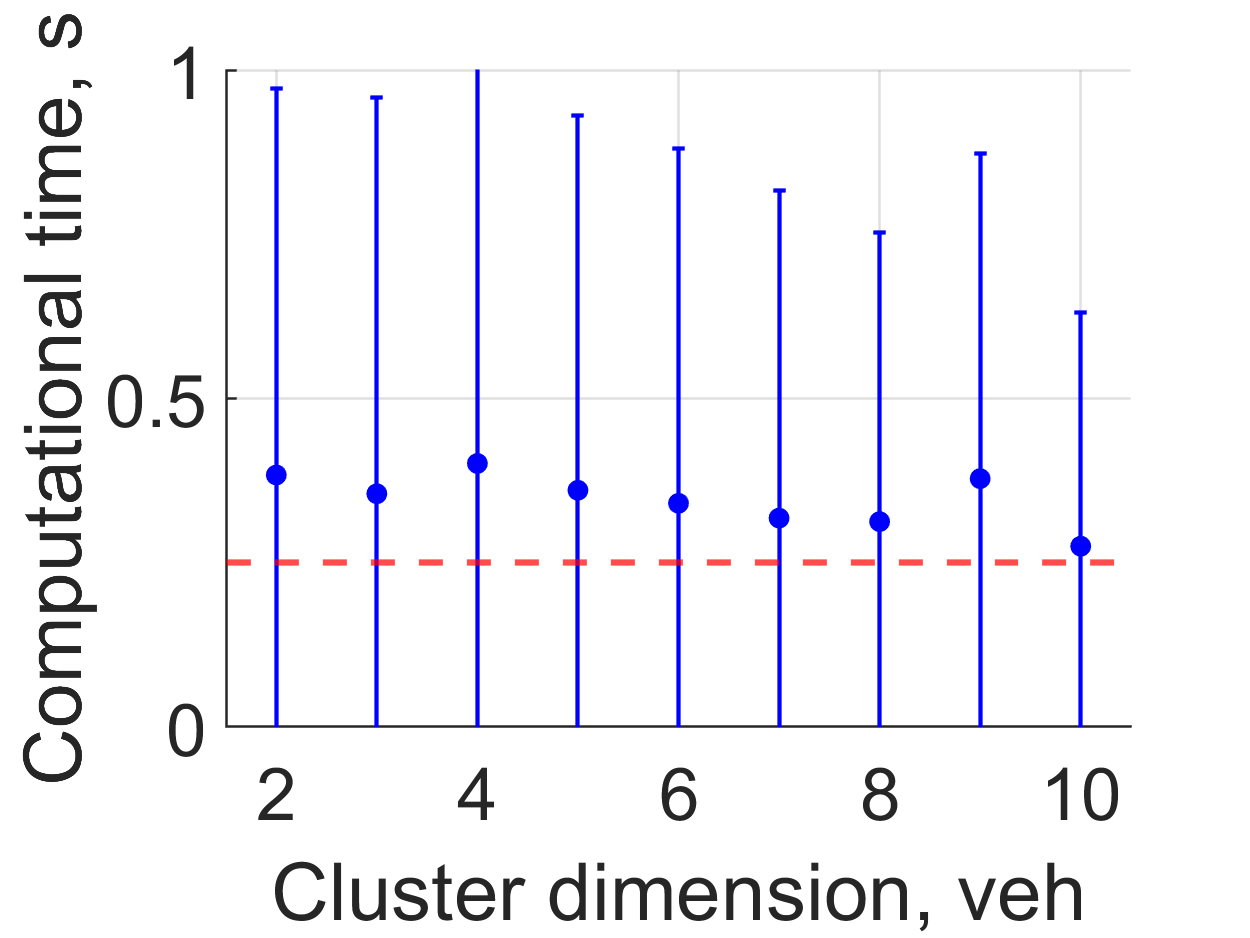}
            \caption{GS\textsubscript{C}}
            \label{fig:ct_gs_c}
        \end{subfigure}
    
        \vspace{4mm} 
    
        \begin{subfigure}[b]{0.48\columnwidth}
            \centering
            \includegraphics[width=\linewidth]{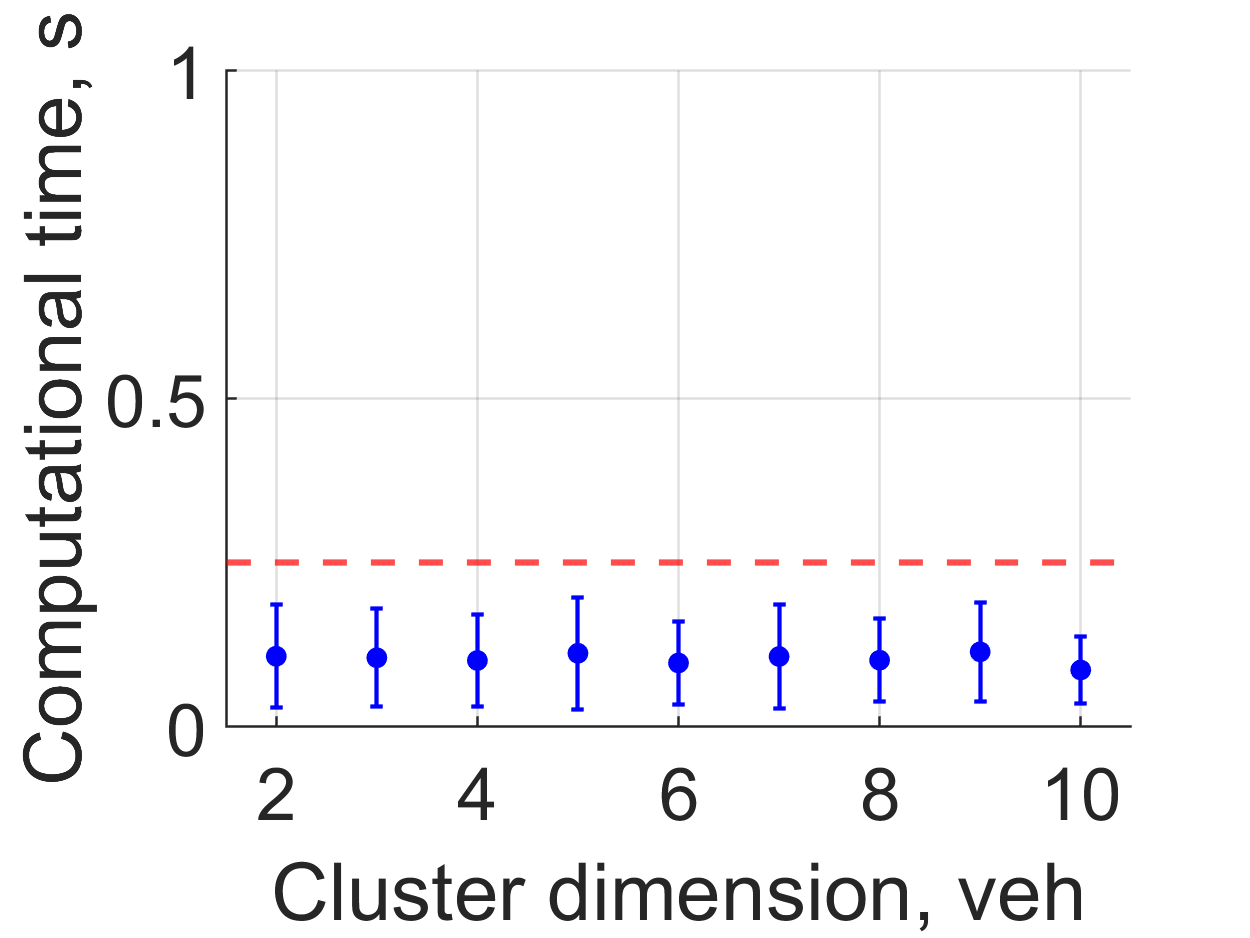}
            \caption{J\textsubscript{D}}
            \label{fig:ct_j_d}
        \end{subfigure}
        \hfill
        \begin{subfigure}[b]{0.48\columnwidth}
            \centering
            \includegraphics[width=\linewidth]{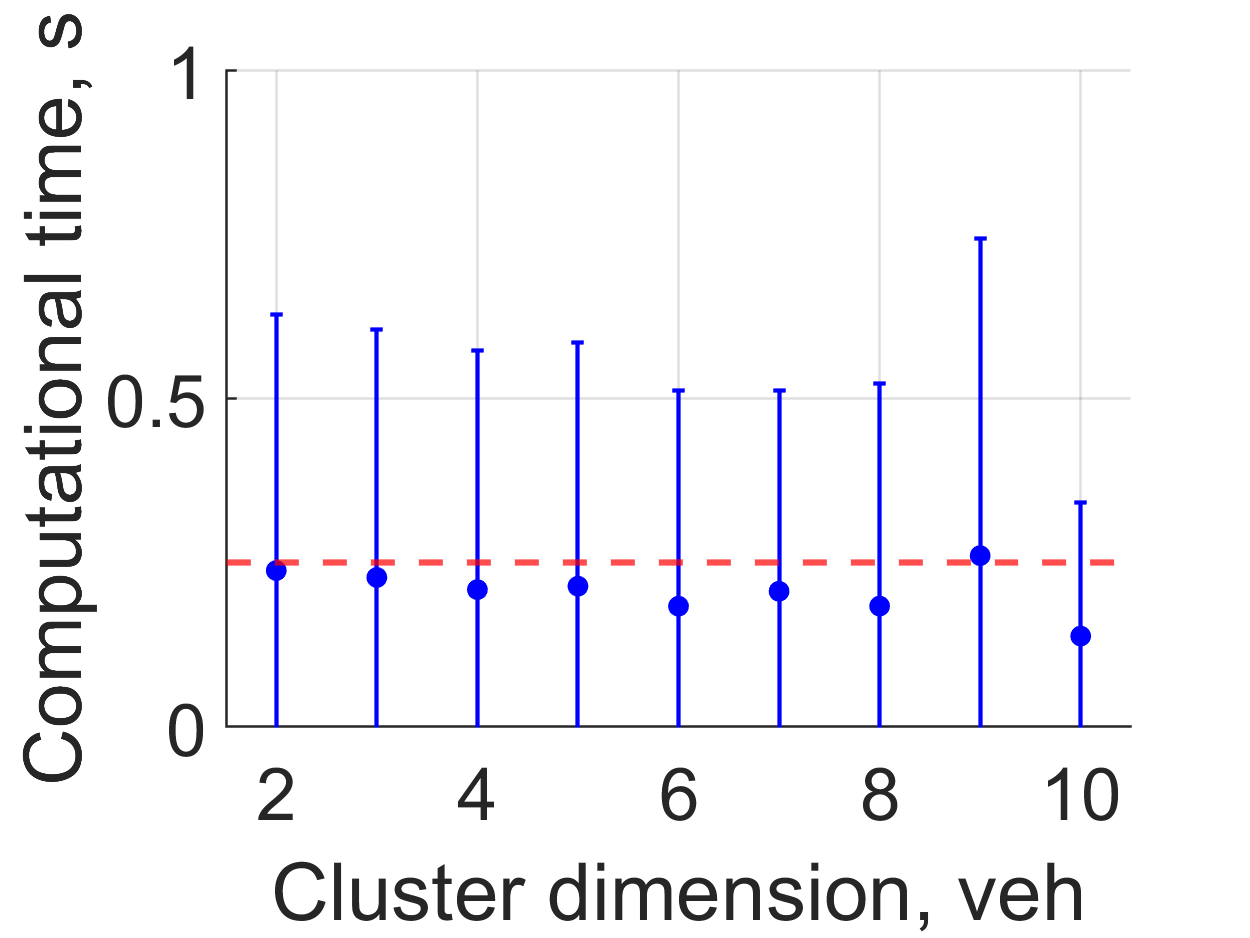}
            \caption{GS\textsubscript{D}}
            \label{fig:ct_gs_d}
        \end{subfigure}
    
        \caption{Computation time vs. cluster size for the decentralized algorithms. Error bars represent standard deviation, and the blue circle is the average. The red line shows the controller sampling time.}
        \label{fig:ct_grid}
    \end{figure}
    The computational performance is evaluated based on the wall-clock time per solver instance relative to the cluster dimension. For this purpose, we analyze all clusters that appear in both congested and uncongested scenarios. As observed in the simulation results, computation time is largely independent of the cluster size, with no significant upward trend as the number of CAVs in a threat cluster increases. 
    
    Comparing the solvers, the Jacobi algorithm consistently outperforms the GS algorithm in terms of speed and lower standard deviation across all cluster dimensions. Similarly, the decoupled variants (J\textsubscript{D}, GS\textsubscript{D}) are faster than their coupled counterparts (J\textsubscript{C}, GS\textsubscript{C}). The fastest configuration is J\textsubscript{D}, where the average computation time, including the upper bound of the standard deviation, remains below the sampling time of 0.25\,s for all cluster sizes. 

    
    \begin{figure*}[t] 
        \centering
        
        \begin{subfigure}{\textwidth}
            \centering
            \includegraphics[width=\textwidth]{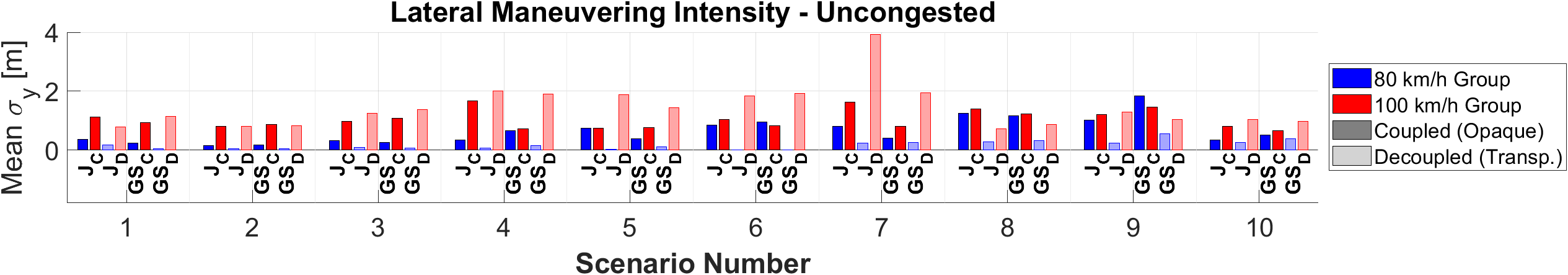}
            \caption{Uncongested}
            \label{fig:uncon_lat}
        \end{subfigure}
        \vspace{1em}
        \begin{subfigure}{\textwidth}
            \centering
            \includegraphics[width=\textwidth]{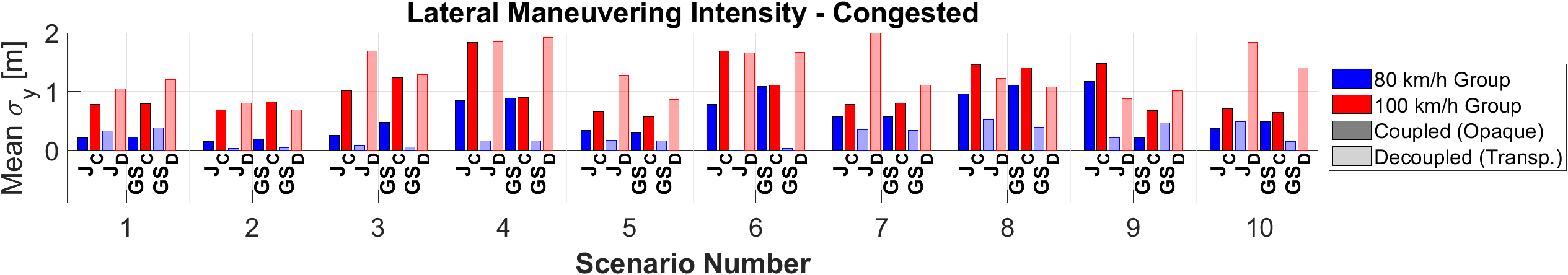}
            \caption{Congested}
            \label{fig:con_lat}
        \end{subfigure}
        
        \caption{Mean of the standard deviation of the lateral position of fast CAVs (red) and slow CAVs (blue) for all scenarios and controllers (J\textsubscript{C}, J\textsubscript{D}, GS\textsubscript{C}, GS\textsubscript{D}).}
        \label{fig:lateral_comparison}
    \end{figure*}
    \subsection{Performance}\label{subsec:perf}
    \begin{table}[t]
        \centering
        \caption{RMSE (mean $\pm$ std) for all methods averaged over all CAVs and scenarios.}
        \label{tab:rmse_perf}
        \setlength{\tabcolsep}{6pt} 
        \renewcommand{\arraystretch}{1.2}
        \begin{tabular}{lccc}
            \hline
            \textbf{Method} & \textbf{Velocity} & \textbf{Steering} & \textbf{Acceleration} \\
            & [m/s] & [$10^{-2}$rad] & [m/s$^2$] \\
            \hline
            J\textsubscript{C}  & $1.16 \pm 0.89$ & $0.94 \pm 0.56$ & $2.44 \pm 1.44$ \\
            J\textsubscript{D}  & $0.70 \pm 1.22$ & $0.91 \pm 1.21$ & $0.87 \pm 0.91$ \\
            GS\textsubscript{C} & $1.21 \pm 1.04$ & $0.85 \pm 0.54$ & $2.38 \pm 1.54$ \\
            GS\textsubscript{D} & $0.50 \pm 0.99$ & $0.80 \pm 0.78$ & $0.70 \pm 0.78$ \\
            \hline
        \end{tabular}
    \end{table}
    The tracking quality of the decentralized controllers is quantified using the Root Mean Square Error (RMSE) for velocity, heading angle, and acceleration (Table \ref{tab:rmse_perf}). These three variables are the ones that are accounted for in the quadratic cost function \ref{eq:cost_func} of OCP \ref{eq:DMPC_OCP}. The weighting matrices $Q$ and $R$ are identical over all methods. Across all simulated scenarios, decentralized variants consistently demonstrate high tracking precision. Among these variants, the decoupled versions (J\textsubscript{D}, GS\textsubscript{D}) achieve lower mean RMSE than their coupled counterparts (J\textsubscript{C}, GS\textsubscript{C}). Specifically, GS\textsubscript{D} exhibits the highest absolute tracking accuracy with a mean velocity RMSE of 0.5 m/s. Furthermore, a clear performance trend is observed: GS variants generally outperform Jacobi variants in tracking accuracy, maintaining lower errors during complex maneuvers involving multiple threats.\\
    Another notable point is the different behavior of the coupled and decoupled versions illustrated by Figure \ref{fig:lateral_comparison}. We consider the lateral movement of the slow ($v_{\text{d}}=$80 km/h) and fast ($v_{\text{d}}=$100 km/h) groups separately. For that, we determine the standard deviation of the lateral position for each CAV and take the mean for the slow and fast groups, respectively ($\bar{\sigma}_{\text{y}}$). Figure \ref{fig:uncon_lat} shows the results for each controller and each scenario for the uncongested case, and Figure \ref{fig:con_lat} shows the results for the congested case. We observe that in the uncongested case, $\bar{\sigma}_{\text{y}}$ of the slow group for the decoupled version is lower than $\bar{\sigma}_{\text{y}}$ of the slow group of the coupled version for all scenarios. This also holds for the congested case, apart from a few exceptions. For the fast group, one can see that in most scenarios (uncongested and congested), $\bar{\sigma}_{\text{y}}$ is similar between the coupled and decoupled versions, with a slight tendency to be higher in the decoupled case. This suggests that the fast group naturally navigates through the given spatial composition of the slow group in the decoupled versions. In the coupled versions, the slow group tries to create space for the fast group. This intuitively manifests the decoupling effect of Algorithm \ref{algo:decoupling}.

    \section{Discussion}
	\label{sec:discussion}
	The simulation results demonstrate that the proposed decentralized control architectures can effectively manage complex lane-free highway traffic.

    \subsection{Safety}
    Safety is the most critical metric for highway applications, and our results in Fig. \ref{fig:safe_ell_rec} provide empirical evidence that all four proposed controllers maintain a high safety standard. Throughout the extensive simulation case studies, no collisions were recorded for any of the control variants. Since the coupled versions (J\textsubscript{C}, GS\textsubscript{C}) use additional soft constraints with a large slack penalty in addition to the necessary hard constraints, their behavior is more conservative than that of the decoupled versions (J\textsubscript{D}, GS\textsubscript{D}) that construct the tangents to the hard constraints as described in Fig. \ref{fig:geometric_decoupling_full}. However, note that the decoupled versions are still more conservative than the coupled versions with zero slack penalty.
    
    \subsection{Scalability and Real-Time Feasibility}
    The computational results indicate that the proposed DMPC schemes offer the scalability required for real-time Intelligent Transportation Systems (ITS) applications. A significant observation is that the computation time remains largely independent of the cluster dimensions for all controllers, exhibiting no clear upward trend as the number of interacting CAVs increases.\\
    In terms of computational cost, J\textsubscript{D} emerges as the most efficient, consistently staying below the controller sampling time. 
    The speed-up in the decoupled versions is attributed to replacing nonconvex safety constraints with simpler convex half-spaces. This transformation reduces the complexity of the underlying Non-Linear Programs (NLPs), requiring lower per-iteration computation time to reach optimality.\\
    Interestingly, the computation times for GS\textsubscript{C} and GS\textsubscript{D} do not scale with the cluster size, despite the algorithm's sequential nature. This suggests that using updated trajectories from threats yields progressively better-conditioned optimization problems. Thus, lower per-iteration computation time is needed, which offsets the sequential delay inherent in the GS structure.
    
    \subsection{Performance}

    The GS variants (Algorithm \ref{algo:gauss_seidel}) achieve superior tracking precision because they allow for intra-iteration information propagation, unlike the Jacobi scheme (Algorithm \ref{algo:jacobi}). By incorporating a threat's updated maneuver immediately into its own optimization, a CAV can adjust more precisely to the planned movements of surrounding threats.\\    
    Furthermore, the improved performance of the decoupled versions suggests that transforming nonconvex safety regions into convex half-spaces facilitates more efficient optimization trajectories, enabling solvers to reach higher-quality solutions within the predefined control time step of 0.25 s.\\
    

    \section{Summary and Future Work}
    \label{sec:summary}
    This paper introduced a DMPC framework for lane-free CAV coordination, utilizing a novel geometric decoupling method to transform non-convex safety constraints into convex half-spaces. Evaluated across Jacobi and Gauss-Seidel solvers, the results confirm that all proposed variants maintain empirical safety. While Gauss-Seidel variants offer superior tracking precision, the Jacobi-Decoupled (J\textsubscript{D}) variant proves most effective for real-time deployment. Ultimately, the transition to decoupled constraints substantially improves numerical stability and efficiency. 
    
    Future work will focus on reducing conservatism in the decoupling step by using consensus-based algorithms, such as ADMM, to enable CAVs to jointly negotiate shared half-space boundaries. This iterative consensus approach aims to recover the performance of centralized coordination while maintaining decentralized scalability. 
	\section*{ACKNOWLEDGMENTS}
	This research was supported as part of NCCR Automation, funded by the Swiss National Science Foundation (grant number 51NF40 225155).

\bibliography{references}

@techreport{chavoshi_fairness_2024,
  author      = {Chavoshi, Kimia and Ferrara, Antonella and Kouvelas, Anastasios},
  title       = {Introducing Fairness in Lane-Free Traffic: The Application of Karma Games to Enforce Fair Collaboration of CAVs},
  institution = {TechRxiv},
  type        = {Preprint},
  year        = {2024},
  month       = {09},
  doi         = {10.36227/techrxiv.172668503.39089769v1}
}

@article{christofides_distributed_2013,
  author  = {Christofides, Panagiotis D. and Scattolini, Riccardo and Mu{\~n}oz de la Pe{\~n}a, David and Liu, Jinfeng},
  title   = {Distributed Model Predictive Control: A Tutorial Review and Future Research Directions},
  journal = {Computers \& Chemical Engineering},
  volume  = {51},
  pages   = {21--41},
  year    = {2013},
  issn    = {0098-1354},
  doi     = {10.1016/j.compchemeng.2012.05.011}
}

@book{rawlings_mpc_2020,
  author    = {Rawlings, James B. and Mayne, David Q. and Diehl, Moritz},
  title     = {Model Predictive Control: Theory and Design},
  edition   = {2},
  publisher = {Nob Hill Publishing},
  address   = {Madison, WI},
  year      = {2020},
  isbn      = {978-0-9759377-2-5}
}

@book{maestre2014dmpc,
  title     = {Distributed Model Predictive Control Made Easy},
  editor    = {Maestre, J. M. and Negenborn, R. R.},
  series    = {Intelligent Systems, Control and Automation: Science and Engineering},
  volume    = {69},
  publisher = {Springer},
  address   = {Dordrecht, The Netherlands},
  year      = {2014}
}

@article{bai2023robust,
  author  = {Bai, Weiqi and Xu, Bin and Liu, Hui and Qin, Yechen and Xiang, Changle},
  title   = {Robust Longitudinal Distributed Model Predictive Control of Connected and Automated Vehicles with Coupled Safety Constraints},
  journal = {IEEE Transactions on Vehicular Technology},
  volume  = {72},
  number  = {3},
  pages   = {2960--2972},
  year    = {2023},
  month   = {03}
}

@article{katriniok2022fully,
  author  = {Katriniok, Alexander and Rosarius, Benedikt and M{\"a}h{\"o}nen, Petri},
  title   = {Fully Distributed Model Predictive Control of Connected Automated Vehicles in Intersections: Theory and Vehicle Experiments},
  journal = {IEEE Transactions on Intelligent Transportation Systems},
  volume  = {23},
  number  = {10},
  pages   = {18288--18300},
  year    = {2022},
  month   = {10},
  doi     = {10.1109/TITS.2022.3162038}
}

@article{dabestani2025distributed,
  author  = {Dabestani, N. and others},
  title   = {Distributed Model Predictive Control of Automated Vehicle Entities in a Lane-Free Environment: An Event-Triggered Approach},
  journal = {Control Engineering Practice},
  volume  = {127},
  pages   = {107839},
  year    = {2025},
  doi     = {10.1016/j.conengprac.2025.107839}
}

@article{pierer2024sensitivity,
  author        = {Pierer von Esch, Maximilian and V{\"o}lz, Andreas and Graichen, Knut},
  title         = {Sensitivity-Based Distributed Model Predictive Control for Nonlinear Systems under Inexact Optimization},
  journal       = {arXiv preprint arXiv:2406.03134},
  year          = {2024},
  eprint        = {arXiv:2406.03134},
  archivePrefix = {arXiv}
}

@article{liu2024admm,
  author        = {Liu, Haichao and Huang, Zhenmin and Zhu, Zicheng and Li, Yulin and Shen, Shaojie and Ma, Jun},
  title         = {Improved Consensus ADMM for Cooperative Motion Planning of Large-Scale Connected Autonomous Vehicles with Limited Communication},
  journal       = {arXiv preprint arXiv:2401.09032},
  year          = {2024},
  eprint        = {arXiv:2401.09032},
  archivePrefix = {arXiv}
}

@article{dong2024eco,
  author  = {Dong, Shiying and Ghezzi, Andrea and others},
  title   = {Real-Time NMPC with Convex--Concave Constraints and Application to Eco-Driving},
  journal = {IEEE Transactions on Control Systems Technology},
  year    = {2024}
}

@inproceedings{levy_path_2021,
  author    = {Levy, R. and Haddad, J.},
  title     = {Path and Trajectory Planning for Autonomous Vehicles on Roads without Lanes},
  booktitle = {Proceedings of the 2021 IEEE International Intelligent Transportation Systems Conference (ITSC)},
  pages     = {3871--3876},
  year      = {2021},
  publisher = {IEEE}
}

@article{doan_jacobi_2008,
  author  = {Doan, Dang and Keviczky, Tam{\'a}s and Necoara, Ion and Diehl, Moritz},
  title   = {A Jacobi Algorithm for Distributed Model Predictive Control of Dynamically Coupled Systems},
  journal = {arXiv preprint arXiv:0809.3647},
  year    = {2008},
  eprint  = {arXiv:0809.3647},
  archivePrefix = {arXiv}
}

@inproceedings{sekeran_lane_free_2022,
  author    = {Sekeran, M. and Rostami-Shahrbabaki, M. and Syed, A. A. and Margreiter, M. and Bogenberger, K.},
  title     = {Lane-Free Traffic: History and State of the Art},
  booktitle = {Proceedings of the 2022 IEEE 25th International Conference on Intelligent Transportation Systems (ITSC)},
  pages     = {1037--1042},
  year      = {2022},
  publisher = {IEEE},
  doi       = {10.1109/ITSC55140.2022.9922309}
}

@inproceedings{facerias_nonlinear_2024,
  author    = {Facer{\'i}as, Marc and Puig, Vicen{\c{c}} and Stancu, Alexandru},
  title     = {Non-Linear Distributed MPC Coordination of Autonomous Vehicles Using Optimality Condition Decomposition},
  booktitle = {Proceedings of the 2024 European Control Conference (ECC)},
  year      = {2024},
  publisher = {IEEE},
  doi       = {10.23919/ECC64448.2024.10591303}
}

@inproceedings{lygerosADMM,
  author    = {Rey, Felix and Pan, Zhoudan and Hauswirth, Adrian and Lygeros, John},
  title     = {Fully Decentralized ADMM for Coordination and Collision Avoidance},
  booktitle = {Proceedings of the 2018 European Control Conference (ECC)},
  pages     = {825--830},
  year      = {2018},
  doi       = {10.23919/ECC.2018.8550245}
}

@incollection{voronoi_graphic,
  author    = {Cuenca Macas, Leduin Jos{\'e} and Pineda, Israel},
  title     = {Collision Avoidance Simulation Using Voronoi Diagrams in a Centralized System of Holonomic Multi-Agents},
  booktitle = {Information and Communication Technologies},
  editor    = {Herrera-Tapia, Jorge and Rodriguez-Morales, Germania and Fonseca C., Efra{\'i}n R. and Berrezueta-Guzman, Santiago},
  pages     = {18--31},
  publisher = {Springer International Publishing},
  address   = {Cham},
  year      = {2022},
  isbn      = {978-3-031-18272-3}
}

@inbook{Papageorgiou2024,
  author    = {Papageorgiou, Markos and others},
  editor    = {Ioannou, Petros and Malikopoulos, Andreas A.},
  title     = {Highlights of Lane-Free Automated Vehicle Traffic with Nudging},
  booktitle = {Transportation Mobility in Smart Cities},
  year      = {2024},
  publisher = {Springer Nature Switzerland},
  address   = {Cham},
  pages     = {147--183},
  doi       = {10.1007/978-3-031-64769-7_6}
}

@article{chavoshi2025,
  title={Integrated internal boundary control and ramp metering in lane-free highway systems: A combined feedback linearization and mpc approach},
  author={Chavoshi, Kimia and Malekzadeh, Milad and Papageorgiou, Markos and Ferrara, Antonella and Kouvelas, Anastasios},
  journal={IEEE Transactions on Intelligent Transportation Systems},
  year={2025},
  publisher={IEEE},
  doi={ 10.1109/TITS.2025.3577459}
}

\end{document}